\documentclass[aps,prx,twocolumn,superscriptaddress,floatfix,longbibliography,10pt,floatfix]{revtex4-2}

\usepackage[utf8]{inputenc}
\usepackage{braket}
\usepackage{amssymb,amsmath,amsbsy,mathtools}
\usepackage{braket}
\usepackage{graphicx}
\usepackage{subfigure}
\usepackage[colorlinks=true,citecolor=blue,linkcolor=black]{hyperref}

\newcommand{\mG}{\mathcal{G}}
\newcommand{\mbM}{\mathbb{M}}
\newcommand{\mbR}{\mathbb{R}}
\newcommand{\mbS}{\mathbb{S}}

\newcommand{\tr}{\text{Tr}}

\begin{document}

\title{Reinforced Disentanglers on Random Unitary Circuits}

\begin{abstract}
    We search for efficient disentanglers on random Clifford circuits of two-qubit gates arranged in a brick-wall pattern using the proximal policy optimization (PPO) algorithm \cite{schulman2017proximalpolicyoptimizationalgorithms}. Disentanglers are a set of projective measurements inserted between consecutive entangling layers. An efficient disentangler is a set of projective measurements that minimize the averaged von Neumann entropy of the final state with the least number of total projections possible. The problem is naturally amenable to reinforcement learning techniques by taking the binary matrix representing the projective measurements along the circuit as our state and actions as bit-flipping operations on this binary matrix that add or delete measurements at specified locations. We give rewards to our agent dependent on the averaged von Neumann entropy of the final state and the configuration of measurements, such that the agent learns the optimal policy that will take him from the initial state of no measurements to the optimal measurement state that minimizes the von Neumann entropy. Our results indicate that the number of measurements required to disentangle a random quantum circuit is drastically less than the numerical results of measurement-induced phase transition papers. Additionally, the reinforcement learning procedure enables us to characterize the pattern of optimal disentanglers, which is not possible in the works of measurement-induced phase transitions.
\end{abstract}

\author{Ning Bao}
\email{ningbao75@gmail.com}
\affiliation{Department of Physics, Northeastern University, Boston, MA, 02115, USA}
\affiliation{Computational Science Initiative, Brookhaven National Laboratory, Upton, NY, 11973, USA}

\author{Keiichiro Furuya}
\email{k.furuya@northeastern.edu}
\affiliation{Department of Physics, Northeastern University, Boston, MA, 02115, USA}

\author{Gün Süer}
\email{suer.g@northeastern.edu}
\affiliation{Department of Physics, Northeastern University, Boston, MA, 02115, USA}

\date{\today}

\maketitle

\tableofcontents
\pagenumbering{gobble}

\section{Introduction}

Measurement-induced phase transition (MIPT) \cite{PhysRevX.7.031016,PhysRevX.9.031009,PhysRevA.62.062311,PhysRevB.99.224307,PhysRevLett.125.030505} is a relatively new class of phase transition observed in random quantum circuits when measurements are incorporated alongside unitary operations. In these circuits, random unitary operations generate entanglement across the system, while measurements tend to collapse quantum states, reducing the overall entanglement. The competition between these processes—entanglement generation and destruction—results in a critical transition. Specifically, as the measurement rate increases, the system undergoes a phase transition from a volume-law phase, where entanglement entropy or von Neumann entropy of a subsystem scales with system size, to an area-law phase, where entanglement entropy scales with the boundary of a subsystem.

The volume-law phase represents a highly entangled quantum state, while in the area-law phase, frequent measurements collapse the quantum state, leading to low entanglement. The transition is non-thermal, emerging from the nature of quantum dynamics rather than thermal fluctuations, making it an example of a non-equilibrium phase transition. This critical point depends on the rate of measurements, and numerical techniques such as tensor networks and Monte Carlo simulations are often used to explore the phase diagram. MIPTs have been explored in various contexts, including random Clifford circuits, stabilizer circuits, and hybrid quantum systems.

These transitions are significant for understanding the limitations of quantum error correction and how information scrambling behaves in quantum systems \cite{breuer2002theory}. Li et al. (2018) \cite{Li_2018} discussed how the quantum Zeno effect plays a role in this transition from volume to area law in random circuits. Skinner et al. (2019) \cite{PhysRevX.9.031009} presented the numerical evidence for the existence of these transitions in entanglement dynamics. Chan et al. (2019) \cite{Chan_2019} extended the understanding of the entanglement dynamics in unitary-projective hybrid circuits. Together, these works lay the theoretical and numerical foundation for studying MIPTs.

Even though great insight has been gained regarding entanglement properties of one-dimensional qubit systems through numerical simulations studying measurement-induced phase transitions, the probabilistic nature of the projections blur the essential properties of the disentanglers. To address that issue, we consider a bottom-up approach for constructing efficient disentanglers in random quantum circuits. In the MIPT literature, brick-wall circuits are composed of consecutive layers of projections and random two-qubit unitaries. In that sense, constructing efficient disentanglers can be framed as a game, in which the disentangler plays against the brick-wall structure, learning how to pick the positions of projections aiming to minimize the reward function consisting of averaged von Neumann entropy of the final state and a configuration of measurements on the circuit.

To solve the disentangling problem, we can take inspiration from reinforcement learning (RL) frameworks \cite{Sutton1998}. In RL, an agent interacts with an environment and learns a policy to maximize a cumulative reward. For our case, the agent (the disentangler) would act within the quantum circuit (the environment), learning optimal strategies for placing projections to minimize the von Neumann entropy. This setup naturally fits the RL paradigm, where actions correspond to the placement of projections, and the reward could be inversely related to the final entropy of the state.

In particular, policy-based methods such as Proximal Policy Optimization (PPO) have shown strong performance in complex, high-dimensional spaces, making them a suitable candidate for this task \cite{schulman2017proximalpolicyoptimizationalgorithms}. PPO operates by iteratively improving the policy based on the expected return while ensuring updates remain within a trust region to avoid overly large policy shifts, which can destabilize learning. The disentangler's task can be framed as learning a policy $\pi_\theta(a|s)$, where $a$ are the actions (placement of projections) and $s$ is the current state of the system, aiming to minimize the entropy after several layers of the quantum circuit.

PPO's key advantage is its balance between exploration and exploitation, allowing the disentangler to efficiently search for better placements without getting stuck in sub-optimal policies. Moreover, PPO's clipped objective function ensures stable learning by penalizing drastic changes in the policy, which is particularly important in stochastic environments like random quantum circuits. Over time, the disentangler would learn a robust policy that can generalize across different configurations of the brick-wall circuit, leading to low-entropy final states with high efficiency.

We organize this letter as follows. In section \ref{sec:methods}, we review stabilizer states and random Clifford structures. Most importantly, the entanglement entropy can be calculated much more efficiently compared to direct diagonalization methods. We also introduce key reinforcement learning concepts and provide the essential features of the PPO algorithm while explaining the reward profile that we used in training efficient disentanglers. In section \ref{sec:results}, we present the result of our numerical simulations. Our results enable the understanding of the pattern for efficient disentanglers, which is currently unattainable with the current random projection models studied in the MIPT literature. We conclude the letter with section \ref{sec:discussions}, where we discuss modifications to the reinforcement learning algorithms and reward functions and comment on possible applications.

\section{Methods}
\label{sec:methods}

We study the dynamics of a 1/2-spin chain with periodic boundary conditions. The time evolution is discrete in the sense that each forward time step corresponds to unitary evolution by a layer of random two-qubit Clifford unitaries. In between each unitary layer, we insert single spin projections along the $z$-component of spin $S_z$. Complementary to the standard MIPT literatures, these measurements are not probabilistic, rather our goal is to learn to optimal set of measurements such that the final state has no entanglement.

\subsection{Random Clifford circuits}

The main tool for characterizing the dynamics is the entanglement entropy. For a pure state $\ket{\psi}$, and a bipartition $(A, \bar{A})$, the $n$-th Renyi entropy is given by
\begin{equation}
    S_n(A) = \frac{1}{1-n} \log \tr (\rho_A)^n,
\end{equation}
where the reduced density matrix for the subsystem $A$ is given by $\rho_A = \tr_{\bar{A}} \ket{\psi}\bra{\psi}$. For general circuits, the computational complexity of the entanglement entropy calculation scales exponentially with the circuit size. However, as Gottesmann and Knill \cite{gottesman1998heisenbergrepresentationquantumcomputers} showed, Clifford circuits (or stabilizer circuits) can still be simulated efficiently even when the entanglement grows exponentially, by encoding the quantum state in terms of stabilizers.

Stabilizer states form the foundations of error correcting codes. For an $n$-qubit system with state $\ket{\psi}$,  its stabilizer group is defined by a complete set of tensored Pauli operators $\mG = \{g_1, g_2, \dots, g_{|\mG|}\}$ such that
\begin{equation}
    g_i \ket{\psi} = +\ket{\psi}.
\end{equation}
The trivial example is the state with all spin-up qubits satisfying
\begin{equation}
    Z_i \ket{\psi} = \ket{\psi}.
\end{equation}

Under the action of a unitary $U$, the state evolves as $\ket{\psi} \to U \ket{\psi}$, while the stabilizer group evolves in the Heisenberg representation
\begin{equation}
    \mG \to \mG^U = \left\{U g_1 U^\dagger, \dots, U g_{|\mG|} U^\dagger \right\}.
\end{equation}
To keep the stabilizer group properties invariant under time evolution, the unitary $U$ must be an element of the Clifford group, such that $g_i^U = U g_i U^\dagger$ is still a tensor product of Pauli operators. Therefore, to simulate a stabilizer state under Clifford circuits, one only needs to keep track of the Pauli strings in the stabilizer group $\mG$, which only takes polynomial time in the number of qubits.

When $\ket{\psi}$ is a code-word, the Renyi entropies are independent of the index $n$, and it's completely determined by the size of the stabilizer group \cite{PhysRevX.7.031016,Hamma_2005_1,Hamma_2005_2}
\begin{equation}
    S(A)  = |A| - \log_2(|\mathcal{G}_A|),
\end{equation}
where $\mG_A$ is a subgroup of $\mG$ that only acts with identity $I$ on qubits that are in $\bar{A}$.

\subsection{Reinforcement learning}

Reinforcement Learning (RL) is a framework in which an agent interacts with an environment, learning a policy $\pi(a|s)$ that maximizes cumulative reward through trial and error. Formulated as a Markov Decision Process (MDP) \cite{puterman1990markov}, the problem is defined by a tuple $(S, A, P, R, \gamma)$, where $S$ is the set of states, $A$ the set of actions, $P(s'|s,a)$ the transition probability between states, $R(s,a)$ the reward function, and $\gamma \in [0, 1]$ the discount factor that balances immediate and future rewards. The agent aims to optimize the expected cumulative reward, expressed as the return $G_t = \sum_{k=0}^{\infty} \gamma^k R_{t+k}$, where $R_t$ is the reward at time $t$. RL methods seek to find an optimal policy $\pi^*$ that maximizes the value function $V^\pi(s) = \mathbb{E}[G_t | s]$, which represents the expected return from state $s$, or its action-value counterpart $Q^\pi(s,a) = \mathbb{E}[G_t | s, a]$.

There are two main categories of RL methods: model-free and model-based. In model-free methods like Q-learning \cite{watkins1992q}, the agent learns directly from the environment without needing a model of state transitions, updating its Q-function iteratively using the Bellman equation
\begin{equation}
    Q(s,a) \leftarrow Q(s,a) + \alpha \left( R + \gamma \max_{a'} Q(s',a') - Q(s,a) \right).
\end{equation}
On the other hand, model-based methods involve learning a model of the environment, often through supervised learning, and using it for planning future actions. Both methods must balance exploration (selecting unfamiliar actions to gather more information) and exploitation (choosing actions that maximize reward based on the current knowledge). Recent advances in deep reinforcement learning (deep RL) combine RL with neural networks, allowing for scalable learning in high-dimensional spaces \cite{mnih2015human}.

\subsubsection{Proximal policy optimization}

Proximal Policy Optimization (PPO) \cite{schulman2017proximalpolicyoptimizationalgorithms} is a reinforcement learning algorithm designed to improve the stability and performance of policy gradient methods while simplifying the computational complexity seen in approaches like Trust Region Policy Optimization (TRPO) \cite{schulman2015trust}. PPO achieves this by limiting the extent of policy updates to prevent overly large updates that could degrade performance. The key idea is to optimize a surrogate objective function that includes a clipped probability ratio, $\frac{\pi_{\theta}(a|s)}{\pi_{\theta_{\text{old}}}(a|s)}$, which ensures that the policy update remains within a specified range, thus avoiding excessively large changes. The objective function for PPO is:
\begin{align}
    L&^{\text{CLIP}}(\theta)\nonumber\\
    &= \mathbb{E}_t \left[\min\left( r_t(\theta) \hat{A}_t, \text{clip}(r_t(\theta), 1-\epsilon, 1+\epsilon)\hat{A}_t \right)\right],
\end{align}
where $r_t(\theta)$ is the probability ratio, $\hat{A}_t$ is the estimated advantage function, and $\epsilon$ is a small hyperparameter controlling the threshold for clipping. This clipping mechanism stabilizes training by preventing large, destructive updates, resulting in improving the performance across a wide range of tasks in continuous and discrete action spaces \cite{schulman2017proximalpolicyoptimizationalgorithms}. PPO has become a popular method in deep reinforcement learning due to its simplicity, ease of implementation, and competitive performance in tasks such as those in the OpenAI Gym \cite{brockman2016openaigym}.

\subsubsection{Rules of the disentangling game}

\begin{figure}
    \centering
    \includegraphics[width=\linewidth]{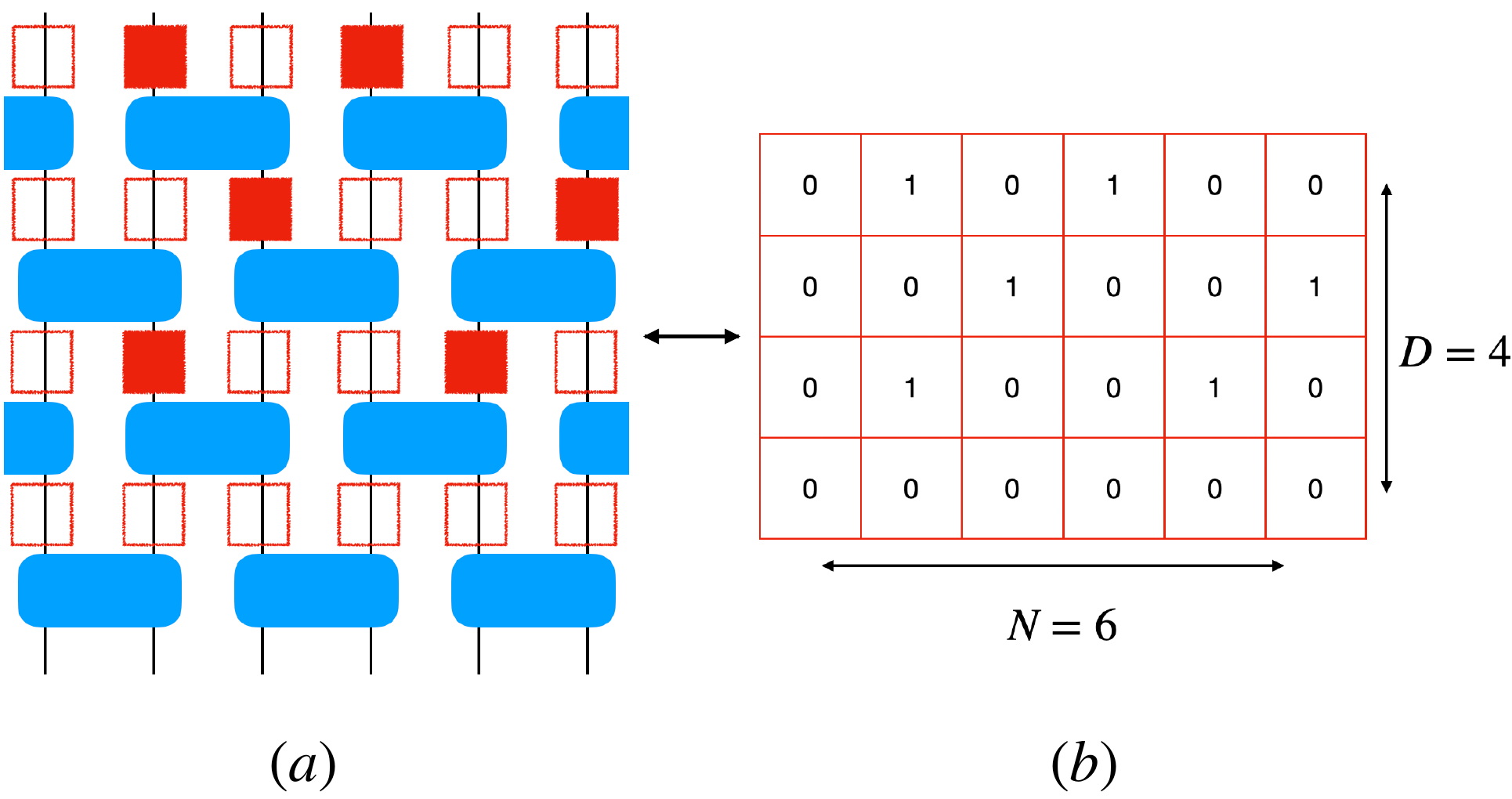}
    \caption{\small{(a) Depiction of the random quantum circuits with the brick-wall structure. Random two-qubit Clifford gates are given in blue, and the projections are given in red (b) The binary matrix that corresponds to the positions of projection operators.}}
    \label{fig:circuit_binarymatrix}
\end{figure}

An environment is prepared by drawing a random Clifford circuit with circuit size $N\times D$. We define the state space and the action space as the set of all possible $N\times D/2$ binary matrices $P$ by translating the length $N$ disentangling circuit layers as length $N$ bit-strings, see Figure \ref{fig:circuit_binarymatrix}. Thus, the binary matrix $P$ is constructed by 
\begin{equation}
    P_{ij} = 
    \begin{cases}
        1 & \; \text{if a measurement is on $(i,j)$}\\
        0 & \; \text{otherwise}.
    \end{cases}
\end{equation}

$P$ starts as a zero matrix at the beginning. At each time step, a single measurement is placed on the circuit corresponding to a single-bit flip in $P$. A single episode is completed when the von Neumann entropy $S_{avg}$ averaged over all the bipartions of a $N$-qubit pure state consisting of a set of subsystems $\{A_1,\cdots,A_N\}$ vanishes, i.e.,
\begin{equation}\label{eq:avg_ee}
    S_{avg} =\frac{1}{N-1} \sum_{i=1}^{N-1}S(A_1, \cdots, A_{i}) = 0.
\end{equation}
This guarantees that the final state is a product state. 

\subsubsection{Rewards and penalties}


We design a sparse reward function. The sparse reward function is evaluated at the end of every episode. 

\begin{figure}
    \centering
    \includegraphics[width=\linewidth]{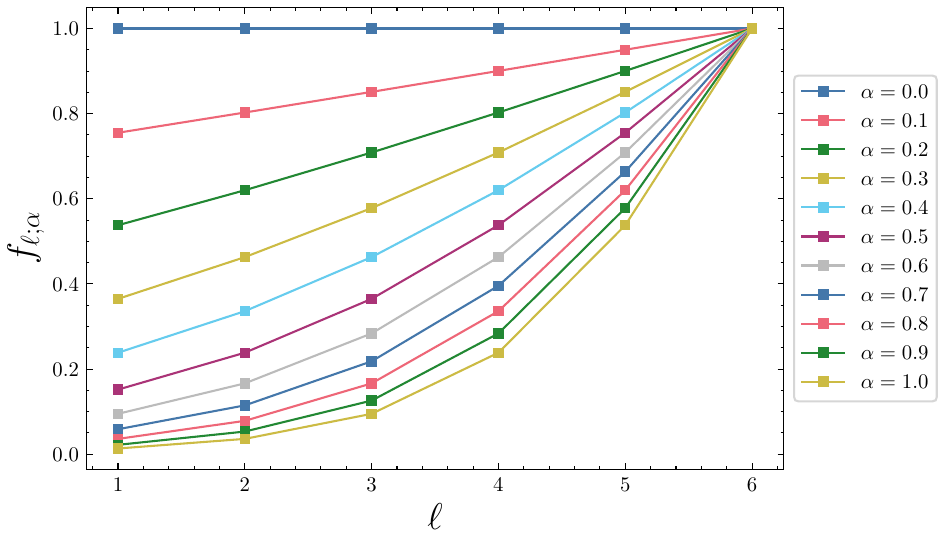}
    \caption{Measurement weights $f_{l; \alpha}$ as a function of layers $l$ for increasing values of penalty slope $\alpha$.}
    \label{fig:penalty_function_alpha}
\end{figure}

Placing the measurement on the last layer at least $N/2$ is the trivial solution. 
To avoid such trivial solutions, both reward functions consider the number of measurements needed to obtain non-trivial efficient disentanglers. Hence, we define a measurement cost as the weighted sum of measurements per layer as
\begin{equation}\label{eq:cost}
    C = \sum_{l=1}^{D/2} f_l m_l
\end{equation}
where $0\leq m_l\leq N $ is the number of measurements at the $l$-th layer. The weights $0\leq f_l\leq 1$ is a monotonically increasing penalty rate as a layer gets deeper, i.e.,
\begin{equation}
    f_l > f_l' \text{ for $l>l'$ }
\end{equation}
where $l,l' \in D/2$ are the labels of $l$-th and $l'$-th layers. We adopt the following weight function,
\begin{equation}\label{eq:penalty}
    f_{l;\alpha} = \frac{2e^{-\alpha l }}{1+e^{-\alpha l }}
\end{equation}
where a penalty slope $\alpha \in \mbR_+$ controls the slope, see figure \ref{fig:penalty_function_alpha}.

Then, the sparse reward function is defined using (\ref{eq:cost}) as
\begin{equation}\label{eq:sparse}
    R= 1- \frac{C}{F N}
\end{equation}
where $F=\sum_l^{D/2} f_{l;\alpha}$, and $0\leq R \leq 1$.


\section{Results}
\label{sec:results}

In this section, we present the results of numerical simulations of reinforced disentanglers. We study the performance of the PPO agent as a function of $N$, $D$, and $\alpha$. The quality and behavior of the learning process are discussed in Appendix \ref{app:num-methods} alongside the plots for the reinforcement learning metrics as a function of time.

We define the averaged number of measurements minimized with respect to a reward function $R$ over the final configurations $P_f$ of measurements  as
\begin{equation}
    \braket{M}_{eps}:= \underset{P_f}{min|_R} \braket{\tilde{M}}_{eps}.
\end{equation}
where $\tilde{M}$ is the number of measurements of the non-optimal configuration. For the analysis below, we further simplify it as 
\begin{equation}
    \mbM := \braket{M}_{eps}.
\end{equation}

In the following subsections, we denote any multi-linear function $\eta(x_1, x_2, \cdots, x_n)$ on variables $x_1, \cdots, x_n$ for some $n$ as 
\begin{equation}
    \eta^{x_i,x_j}(x_1,\cdots, x_{i-1}, x_{i+1}, \cdots,x_{j-1}, x_{j+1}, \cdots, x_n)
\end{equation}
when evaluating $\eta$ for a fixed $x_i$ and $x_j$, for instance.


\subsection{Sparse rewards}

\begin{figure}
    \centering
    \includegraphics[width=\linewidth]{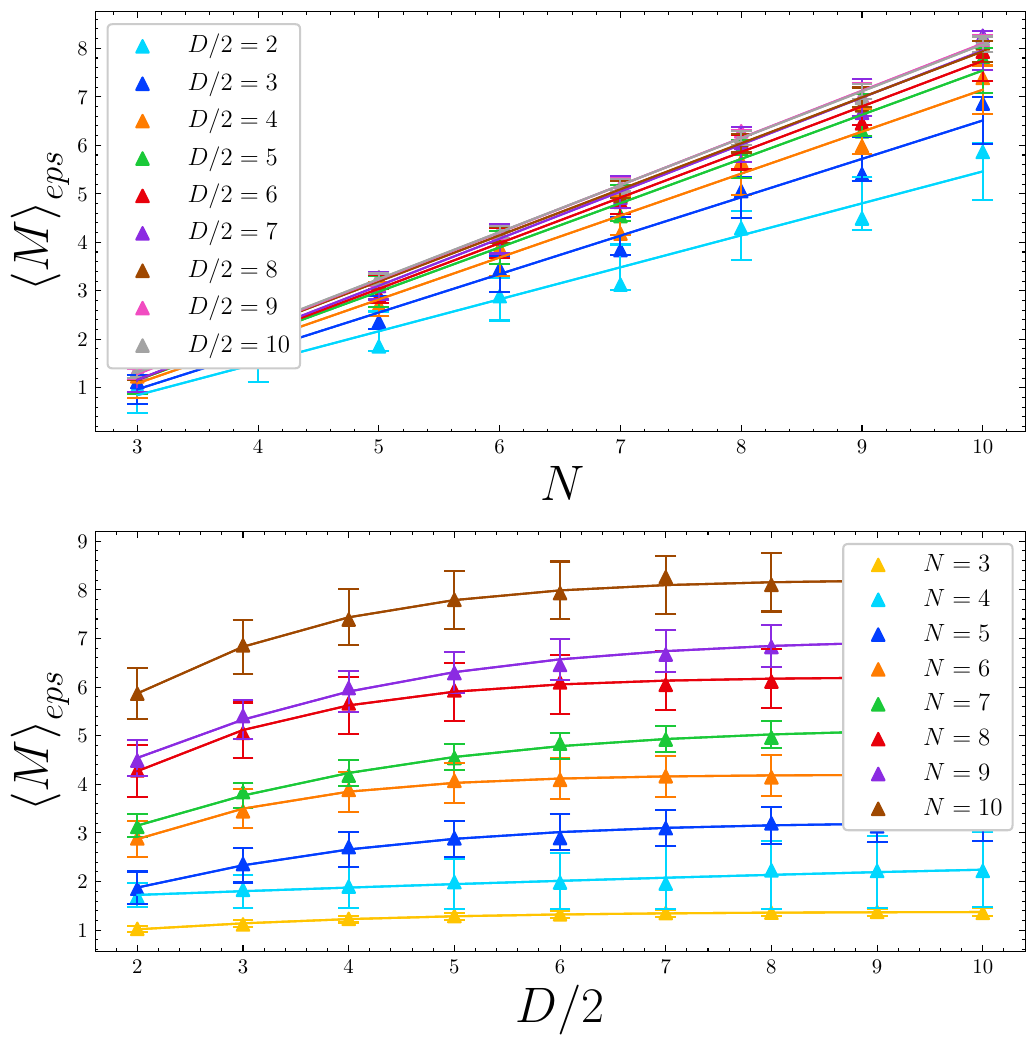}
    \caption{\small{Rewards and number of projections averaged over episodes vs the number of qubits the trained PPO models, with $\alpha=0.1$, $t_s = 250,000$, $\ell_r = 0.1$, $e_c = 0.01$, and positive sparse reward $p_r = 50.0$. The best fit of the form (Top) $y = \gamma_1 \tanh(\gamma_2 x) + \gamma_3$ and (Bottom) $y = \gamma_1 x + \gamma_2 $ are displayed with error bars. (Bottom) Fit parameters can be found in table \ref{tab:my_label}.}}
    \label{fig:aa_01_len_vs_qubits_depth}
\end{figure}



When the sparse reward function \eqref{eq:sparse} is adopted, the number of disentanglers averaged over episodes $\mbM(N,D,\alpha)$ is a function of the number of qubits $N$, the circuit depth $D$, and the penalty slope $\alpha$ in \eqref{eq:penalty}. 

For a fixed depth $D$ and a penalty slope $\alpha$, $\mbM^{D,\alpha}(N)$ depends linearly on the number of qubits $N$ as in figure \ref{fig:aa_01_len_vs_qubits_depth}, i.e., fitted as
\begin{equation}
	\mbM^{D,\alpha}(N)= \gamma_1^{N,\alpha}(D) N + \gamma_2^{N,\alpha}(D)
\end{equation}
where $\gamma_1^N(D,\alpha)$ and $\gamma_2^N(D,\alpha)$ are the fitting parameters as the functions of $D$ and $\alpha$ for a given $N$. Physically, one needs one more measurement on average when two qubits are added\footnote{Odd and even, basically the same. On average, one needs at least $0.5$ more measurement when the number of qubits increased by one.}.

\begin{figure}
    \centering
    \includegraphics[width=\linewidth]{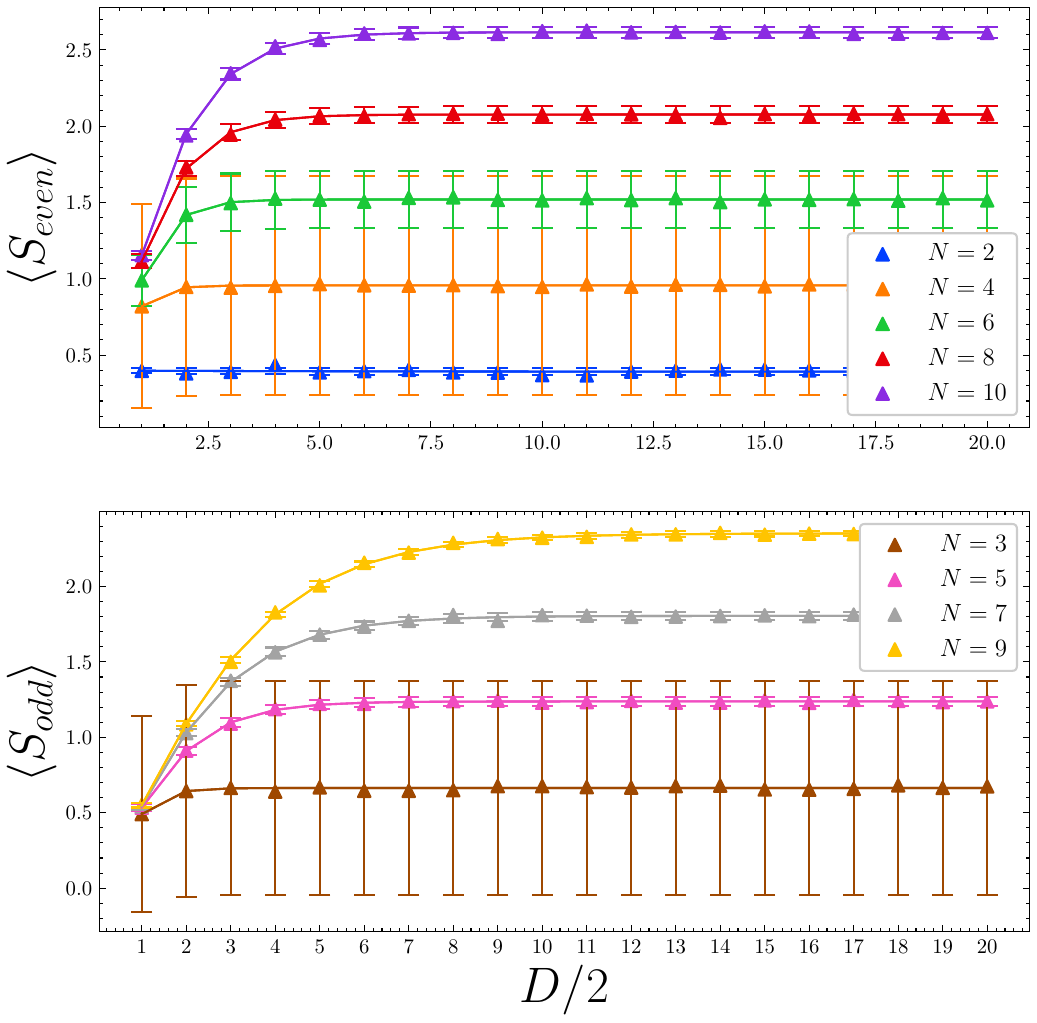}
    \caption{Entanglement growth as a function of depth for brick-wall random Clifford circuits with no projections, with the increasing number of qubits.}
    \label{fig:entanglement_growth}
\end{figure}

For a fixed $N$ and $\alpha$, $\mbM^{N,\alpha}(D)$ behaves as a hyperbolic tangent function on the circuit depth $D$, figure \ref{fig:aa_01_len_vs_qubits_depth}, i.e.,  fitted as
\begin{equation}
	\mbM^{N,\alpha}(D) = \gamma_1^{D}(N,\alpha) \tanh(\gamma_2^{D}(N,\alpha) D)+\gamma_3^{D}(N,\alpha)
\end{equation}
Its monotonically increasing behavior, i.e.,
\begin{equation}
    \mbM^{N,\alpha}(D) \leq \mbM^{N,\alpha}(D')
\end{equation}
for $D<D'$, can be argued based on the numerical fact that 
 \begin{equation}
    \mbS^{N,\alpha}(D) \leq \mbS^{N,\alpha}(D')
\end{equation}
where $\mbS:= \braket{S_{avg}}$ is the averaged von Neumann entropy in \eqref{eq:avg_ee} averaged over $1000$ random circuits, as in Figure \ref{fig:entanglement_growth}
. When the circuit gets deeper, and the averaged von Neumann entropy $\mbS$ in \eqref{eq:avg_ee} stays the same or increases, one needs to increase the number of measurements at least one to keep the final state to be a product state. $\mbM^{N,\alpha}(D)$ saturates to a constant because there need no more additional measurements after the averaged von Neumann entropy $\mbS^{N,\alpha}(D)$ saturates to a constant\footnote{
We leave detailed analytical studies of the behavior from the aspect of entanglement membranes and the percolation theory for future work.}.

\begin{figure}
    \centering
    \includegraphics[width=\linewidth]{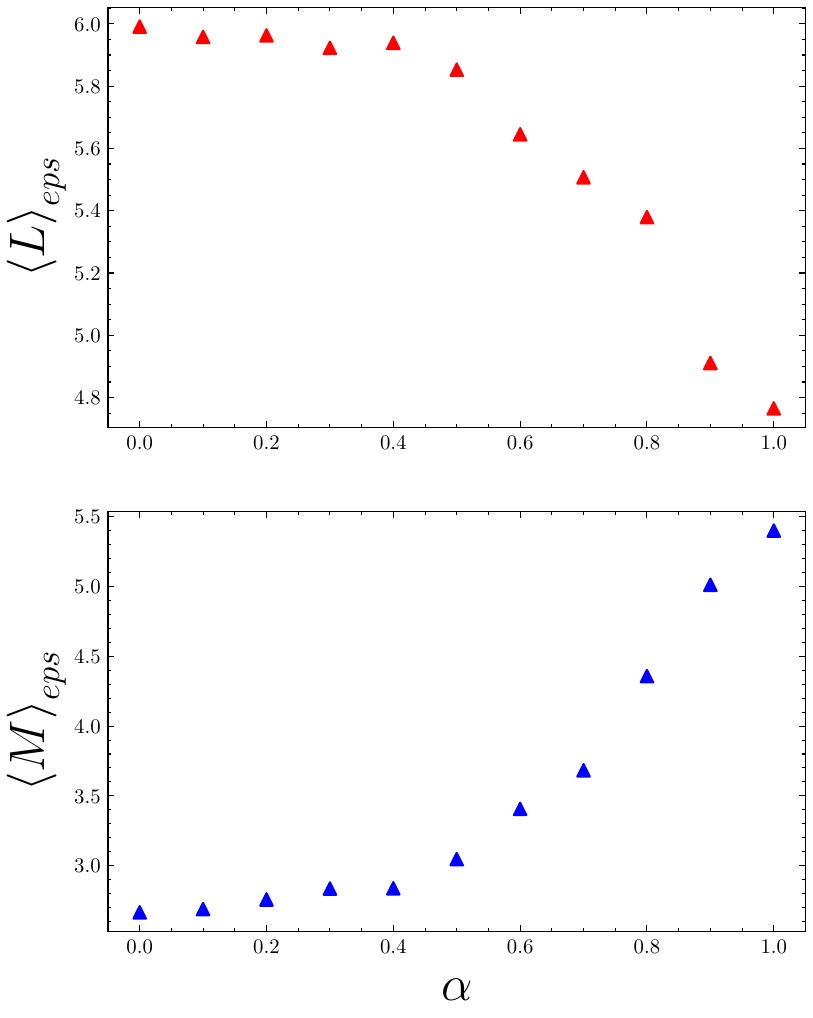}
    \caption{Layers averaged over measurements, and the total number of measurements averaged over 1000 episodes as a function of the penalty slope $\alpha$, for circuits of size $N \times D/2 = 6\times 6$.}
    \label{fig:6x6_alpha_dependence}
\end{figure}

For a fixed number of qubits $N$ and the circuit depth $D$, $\mbM^{N,D}(\alpha)$ increases as in figure \ref{fig:6x6_alpha_dependence}. The measurements on the last layer can contribute to the disentanglement much more than the ones on the earlier layers. 
However, the larger $\alpha$ is, the lesser the number of measurements placed on the later layers is. This results in the increment of $\mbM^{N,D}(\alpha)$ as a function of $\alpha$. In other words, most measurements are on the last layer when $\alpha=0$. As $\alpha$ increases, the measurements tend to be placed away from the later layers. We visualize the phenomenon by defining the weighted average of the layers,
\begin{equation}
    L = \frac{1}{M}\sum_{l}^{D/2 } l m_l
\end{equation}
for a fixed $N$ and $D$, where each $l$-th layer is weighted by the number of measurements on its layer. We plot $L$ averaged over $1000$ episodes $\braket{L}_{eps}$ for $N\times D/2 = 6\times 6$ as a function of $\alpha$ in figure \ref{fig:6x6_alpha_dependence}.



\section{Discussions}
\label{sec:discussions}

Using proximal policy optimization, we have provided a bottom-up approach to understanding the pattern of disentanglers in brick-wall random Clifford circuits. Complementary to the usual direction in the MIPT literature, we try to learn the optimal set of measurements that disentangle random Clifford circuits rather than probabilistically placing them between entangling layers with a fixed rate. This enables a deeper understanding of entanglement structure in one-dimensional spin chains. In future work, we hope to report on modified reinforcement learning schemes and reward functions, as well as implications for multipartite entanglement classification. We would also investigate how reinforcement learning techniques similar to the ones used in this work can be extended to studying quantum resource theories.

\section*{Acknowledgements}
We would like to thank Fabian Ruehle, James Halverson, and Yuki Kagaya for their valuable discussions. N.B. is funded by the Spin Chain Bootstrap for Quantum Computation project from the DOE Office of Science-Basic Energy Sciences, project number PM602. K. F. and G. S. are supported by N.B.'s startup grant at Northeastern University. This work was completed in part using the Discovery cluster, supported by Northeastern University’s Research Computing team.

\bibliographystyle{apsrev4-2}
\bibliography{ref.bib}

\begin{thebibliography}{22}%
\makeatletter
\providecommand \@ifxundefined [1]{%
 \@ifx{#1\undefined}
}%
\providecommand \@ifnum [1]{%
 \ifnum #1\expandafter \@firstoftwo
 \else \expandafter \@secondoftwo
 \fi
}%
\providecommand \@ifx [1]{%
 \ifx #1\expandafter \@firstoftwo
 \else \expandafter \@secondoftwo
 \fi
}%
\providecommand \natexlab [1]{#1}%
\providecommand \enquote  [1]{``#1''}%
\providecommand \bibnamefont  [1]{#1}%
\providecommand \bibfnamefont [1]{#1}%
\providecommand \citenamefont [1]{#1}%
\providecommand \href@noop [0]{\@secondoftwo}%
\providecommand \href [0]{\begingroup \@sanitize@url \@href}%
\providecommand \@href[1]{\@@startlink{#1}\@@href}%
\providecommand \@@href[1]{\endgroup#1\@@endlink}%
\providecommand \@sanitize@url [0]{\catcode `\\12\catcode `\$12\catcode `\&12\catcode `\#12\catcode `\^12\catcode `\_12\catcode `\%12\relax}%
\providecommand \@@startlink[1]{}%
\providecommand \@@endlink[0]{}%
\providecommand \url  [0]{\begingroup\@sanitize@url \@url }%
\providecommand \@url [1]{\endgroup\@href {#1}{\urlprefix }}%
\providecommand \urlprefix  [0]{URL }%
\providecommand \Eprint [0]{\href }%
\providecommand \doibase [0]{https://doi.org/}%
\providecommand \selectlanguage [0]{\@gobble}%
\providecommand \bibinfo  [0]{\@secondoftwo}%
\providecommand \bibfield  [0]{\@secondoftwo}%
\providecommand \translation [1]{[#1]}%
\providecommand \BibitemOpen [0]{}%
\providecommand \bibitemStop [0]{}%
\providecommand \bibitemNoStop [0]{.\EOS\space}%
\providecommand \EOS [0]{\spacefactor3000\relax}%
\providecommand \BibitemShut  [1]{\csname bibitem#1\endcsname}%
\let\auto@bib@innerbib\@empty
\bibitem [{\citenamefont {Schulman}\ \emph {et~al.}(2017)\citenamefont {Schulman}, \citenamefont {Wolski}, \citenamefont {Dhariwal}, \citenamefont {Radford},\ and\ \citenamefont {Klimov}}]{schulman2017proximalpolicyoptimizationalgorithms}%
  \BibitemOpen
  \bibfield  {author} {\bibinfo {author} {\bibfnamefont {J.}~\bibnamefont {Schulman}}, \bibinfo {author} {\bibfnamefont {F.}~\bibnamefont {Wolski}}, \bibinfo {author} {\bibfnamefont {P.}~\bibnamefont {Dhariwal}}, \bibinfo {author} {\bibfnamefont {A.}~\bibnamefont {Radford}},\ and\ \bibinfo {author} {\bibfnamefont {O.}~\bibnamefont {Klimov}},\ }\href {https://arxiv.org/abs/1707.06347} {\bibinfo {title} {Proximal policy optimization algorithms}} (\bibinfo {year} {2017}),\ \Eprint {https://arxiv.org/abs/1707.06347} {arXiv:1707.06347 [cs.LG]} \BibitemShut {NoStop}%
\bibitem [{\citenamefont {Nahum}\ \emph {et~al.}(2017)\citenamefont {Nahum}, \citenamefont {Ruhman}, \citenamefont {Vijay},\ and\ \citenamefont {Haah}}]{PhysRevX.7.031016}%
  \BibitemOpen
  \bibfield  {author} {\bibinfo {author} {\bibfnamefont {A.}~\bibnamefont {Nahum}}, \bibinfo {author} {\bibfnamefont {J.}~\bibnamefont {Ruhman}}, \bibinfo {author} {\bibfnamefont {S.}~\bibnamefont {Vijay}},\ and\ \bibinfo {author} {\bibfnamefont {J.}~\bibnamefont {Haah}},\ }\href {https://doi.org/10.1103/PhysRevX.7.031016} {\bibfield  {journal} {\bibinfo  {journal} {Phys. Rev. X}\ }\textbf {\bibinfo {volume} {7}},\ \bibinfo {pages} {031016} (\bibinfo {year} {2017})}\BibitemShut {NoStop}%
\bibitem [{\citenamefont {Skinner}\ \emph {et~al.}(2019)\citenamefont {Skinner}, \citenamefont {Ruhman},\ and\ \citenamefont {Nahum}}]{PhysRevX.9.031009}%
  \BibitemOpen
  \bibfield  {author} {\bibinfo {author} {\bibfnamefont {B.}~\bibnamefont {Skinner}}, \bibinfo {author} {\bibfnamefont {J.}~\bibnamefont {Ruhman}},\ and\ \bibinfo {author} {\bibfnamefont {A.}~\bibnamefont {Nahum}},\ }\href {https://doi.org/10.1103/PhysRevX.9.031009} {\bibfield  {journal} {\bibinfo  {journal} {Phys. Rev. X}\ }\textbf {\bibinfo {volume} {9}},\ \bibinfo {pages} {031009} (\bibinfo {year} {2019})}\BibitemShut {NoStop}%
\bibitem [{\citenamefont {Aharonov}(2000)}]{PhysRevA.62.062311}%
  \BibitemOpen
  \bibfield  {author} {\bibinfo {author} {\bibfnamefont {D.}~\bibnamefont {Aharonov}},\ }\href {https://doi.org/10.1103/PhysRevA.62.062311} {\bibfield  {journal} {\bibinfo  {journal} {Phys. Rev. A}\ }\textbf {\bibinfo {volume} {62}},\ \bibinfo {pages} {062311} (\bibinfo {year} {2000})}\BibitemShut {NoStop}%
\bibitem [{\citenamefont {Chan}\ \emph {et~al.}(2019{\natexlab{a}})\citenamefont {Chan}, \citenamefont {Nandkishore}, \citenamefont {Pretko},\ and\ \citenamefont {Smith}}]{PhysRevB.99.224307}%
  \BibitemOpen
  \bibfield  {author} {\bibinfo {author} {\bibfnamefont {A.}~\bibnamefont {Chan}}, \bibinfo {author} {\bibfnamefont {R.~M.}\ \bibnamefont {Nandkishore}}, \bibinfo {author} {\bibfnamefont {M.}~\bibnamefont {Pretko}},\ and\ \bibinfo {author} {\bibfnamefont {G.}~\bibnamefont {Smith}},\ }\href {https://doi.org/10.1103/PhysRevB.99.224307} {\bibfield  {journal} {\bibinfo  {journal} {Phys. Rev. B}\ }\textbf {\bibinfo {volume} {99}},\ \bibinfo {pages} {224307} (\bibinfo {year} {2019}{\natexlab{a}})}\BibitemShut {NoStop}%
\bibitem [{\citenamefont {Choi}\ \emph {et~al.}(2020)\citenamefont {Choi}, \citenamefont {Bao}, \citenamefont {Qi},\ and\ \citenamefont {Altman}}]{PhysRevLett.125.030505}%
  \BibitemOpen
  \bibfield  {author} {\bibinfo {author} {\bibfnamefont {S.}~\bibnamefont {Choi}}, \bibinfo {author} {\bibfnamefont {Y.}~\bibnamefont {Bao}}, \bibinfo {author} {\bibfnamefont {X.-L.}\ \bibnamefont {Qi}},\ and\ \bibinfo {author} {\bibfnamefont {E.}~\bibnamefont {Altman}},\ }\href {https://doi.org/10.1103/PhysRevLett.125.030505} {\bibfield  {journal} {\bibinfo  {journal} {Phys. Rev. Lett.}\ }\textbf {\bibinfo {volume} {125}},\ \bibinfo {pages} {030505} (\bibinfo {year} {2020})}\BibitemShut {NoStop}%
\bibitem [{\citenamefont {Breuer}\ and\ \citenamefont {Petruccione}(2002)}]{breuer2002theory}%
  \BibitemOpen
  \bibfield  {author} {\bibinfo {author} {\bibfnamefont {H.-P.}\ \bibnamefont {Breuer}}\ and\ \bibinfo {author} {\bibfnamefont {F.}~\bibnamefont {Petruccione}},\ }\href@noop {} {\emph {\bibinfo {title} {The theory of open quantum systems}}}\ (\bibinfo  {publisher} {Oxford University Press, USA},\ \bibinfo {year} {2002})\BibitemShut {NoStop}%
\bibitem [{\citenamefont {Li}\ \emph {et~al.}(2018)\citenamefont {Li}, \citenamefont {Chen},\ and\ \citenamefont {Fisher}}]{Li_2018}%
  \BibitemOpen
  \bibfield  {author} {\bibinfo {author} {\bibfnamefont {Y.}~\bibnamefont {Li}}, \bibinfo {author} {\bibfnamefont {X.}~\bibnamefont {Chen}},\ and\ \bibinfo {author} {\bibfnamefont {M.~P.~A.}\ \bibnamefont {Fisher}},\ }\bibfield  {journal} {\bibinfo  {journal} {Physical Review B}\ }\textbf {\bibinfo {volume} {98}},\ \href {https://doi.org/10.1103/physrevb.98.205136} {10.1103/physrevb.98.205136} (\bibinfo {year} {2018})\BibitemShut {NoStop}%
\bibitem [{\citenamefont {Chan}\ \emph {et~al.}(2019{\natexlab{b}})\citenamefont {Chan}, \citenamefont {Nandkishore}, \citenamefont {Pretko},\ and\ \citenamefont {Smith}}]{Chan_2019}%
  \BibitemOpen
  \bibfield  {author} {\bibinfo {author} {\bibfnamefont {A.}~\bibnamefont {Chan}}, \bibinfo {author} {\bibfnamefont {R.~M.}\ \bibnamefont {Nandkishore}}, \bibinfo {author} {\bibfnamefont {M.}~\bibnamefont {Pretko}},\ and\ \bibinfo {author} {\bibfnamefont {G.}~\bibnamefont {Smith}},\ }\bibfield  {journal} {\bibinfo  {journal} {Physical Review B}\ }\textbf {\bibinfo {volume} {99}},\ \href {https://doi.org/10.1103/physrevb.99.224307} {10.1103/physrevb.99.224307} (\bibinfo {year} {2019}{\natexlab{b}})\BibitemShut {NoStop}%
\bibitem [{\citenamefont {Sutton}\ and\ \citenamefont {Barto}(2018)}]{Sutton1998}%
  \BibitemOpen
  \bibfield  {author} {\bibinfo {author} {\bibfnamefont {R.~S.}\ \bibnamefont {Sutton}}\ and\ \bibinfo {author} {\bibfnamefont {A.~G.}\ \bibnamefont {Barto}},\ }\href {http://incompleteideas.net/book/the-book-2nd.html} {\emph {\bibinfo {title} {Reinforcement Learning: An Introduction}}},\ \bibinfo {edition} {2nd}\ ed.\ (\bibinfo  {publisher} {The MIT Press},\ \bibinfo {year} {2018})\BibitemShut {NoStop}%
\bibitem [{\citenamefont {Gottesman}(1998)}]{gottesman1998heisenbergrepresentationquantumcomputers}%
  \BibitemOpen
  \bibfield  {author} {\bibinfo {author} {\bibfnamefont {D.}~\bibnamefont {Gottesman}},\ }\href {https://arxiv.org/abs/quant-ph/9807006} {\bibinfo {title} {The heisenberg representation of quantum computers}} (\bibinfo {year} {1998}),\ \Eprint {https://arxiv.org/abs/quant-ph/9807006} {arXiv:quant-ph/9807006 [quant-ph]} \BibitemShut {NoStop}%
\bibitem [{\citenamefont {Hamma}\ \emph {et~al.}(2005{\natexlab{a}})\citenamefont {Hamma}, \citenamefont {Ionicioiu},\ and\ \citenamefont {Zanardi}}]{Hamma_2005_1}%
  \BibitemOpen
  \bibfield  {author} {\bibinfo {author} {\bibfnamefont {A.}~\bibnamefont {Hamma}}, \bibinfo {author} {\bibfnamefont {R.}~\bibnamefont {Ionicioiu}},\ and\ \bibinfo {author} {\bibfnamefont {P.}~\bibnamefont {Zanardi}},\ }\bibfield  {journal} {\bibinfo  {journal} {Physical Review A}\ }\textbf {\bibinfo {volume} {71}},\ \href {https://doi.org/10.1103/physreva.71.022315} {10.1103/physreva.71.022315} (\bibinfo {year} {2005}{\natexlab{a}})\BibitemShut {NoStop}%
\bibitem [{\citenamefont {Hamma}\ \emph {et~al.}(2005{\natexlab{b}})\citenamefont {Hamma}, \citenamefont {Ionicioiu},\ and\ \citenamefont {Zanardi}}]{Hamma_2005_2}%
  \BibitemOpen
  \bibfield  {author} {\bibinfo {author} {\bibfnamefont {A.}~\bibnamefont {Hamma}}, \bibinfo {author} {\bibfnamefont {R.}~\bibnamefont {Ionicioiu}},\ and\ \bibinfo {author} {\bibfnamefont {P.}~\bibnamefont {Zanardi}},\ }\href {https://doi.org/10.1016/j.physleta.2005.01.060} {\bibfield  {journal} {\bibinfo  {journal} {Physics Letters A}\ }\textbf {\bibinfo {volume} {337}},\ \bibinfo {pages} {22–28} (\bibinfo {year} {2005}{\natexlab{b}})}\BibitemShut {NoStop}%
\bibitem [{\citenamefont {Puterman}(1990)}]{puterman1990markov}%
  \BibitemOpen
  \bibfield  {author} {\bibinfo {author} {\bibfnamefont {M.~L.}\ \bibnamefont {Puterman}},\ }\href@noop {} {\bibfield  {journal} {\bibinfo  {journal} {Handbooks in operations research and management science}\ }\textbf {\bibinfo {volume} {2}},\ \bibinfo {pages} {331} (\bibinfo {year} {1990})}\BibitemShut {NoStop}%
\bibitem [{\citenamefont {Watkins}\ and\ \citenamefont {Dayan}(1992)}]{watkins1992q}%
  \BibitemOpen
  \bibfield  {author} {\bibinfo {author} {\bibfnamefont {C.}~\bibnamefont {Watkins}}\ and\ \bibinfo {author} {\bibfnamefont {P.}~\bibnamefont {Dayan}},\ }\href {https://doi.org/10.1007/BF00992698} {\bibfield  {journal} {\bibinfo  {journal} {Machine Learning}\ }\textbf {\bibinfo {volume} {8}},\ \bibinfo {pages} {279} (\bibinfo {year} {1992})}\BibitemShut {NoStop}%
\bibitem [{\citenamefont {Mnih}\ \emph {et~al.}(2015)\citenamefont {Mnih}, \citenamefont {Kavukcuoglu}, \citenamefont {Silver}, \citenamefont {Rusu}, \citenamefont {Veness}, \citenamefont {Bellemare}, \citenamefont {Graves}, \citenamefont {Riedmiller}, \citenamefont {Fidjeland}, \citenamefont {Ostrovski}, \citenamefont {Petersen}, \citenamefont {Beattie}, \citenamefont {Sadik}, \citenamefont {Antonoglou}, \citenamefont {King}, \citenamefont {Kumaran}, \citenamefont {Wierstra}, \citenamefont {Legg},\ and\ \citenamefont {Hassabis}}]{mnih2015human}%
  \BibitemOpen
  \bibfield  {author} {\bibinfo {author} {\bibfnamefont {V.}~\bibnamefont {Mnih}}, \bibinfo {author} {\bibfnamefont {K.}~\bibnamefont {Kavukcuoglu}}, \bibinfo {author} {\bibfnamefont {D.}~\bibnamefont {Silver}}, \bibinfo {author} {\bibfnamefont {A.~A.}\ \bibnamefont {Rusu}}, \bibinfo {author} {\bibfnamefont {J.}~\bibnamefont {Veness}}, \bibinfo {author} {\bibfnamefont {M.~G.}\ \bibnamefont {Bellemare}}, \bibinfo {author} {\bibfnamefont {A.}~\bibnamefont {Graves}}, \bibinfo {author} {\bibfnamefont {M.}~\bibnamefont {Riedmiller}}, \bibinfo {author} {\bibfnamefont {A.~K.}\ \bibnamefont {Fidjeland}}, \bibinfo {author} {\bibfnamefont {G.}~\bibnamefont {Ostrovski}}, \bibinfo {author} {\bibfnamefont {S.}~\bibnamefont {Petersen}}, \bibinfo {author} {\bibfnamefont {C.}~\bibnamefont {Beattie}}, \bibinfo {author} {\bibfnamefont {A.}~\bibnamefont {Sadik}}, \bibinfo {author} {\bibfnamefont {I.}~\bibnamefont {Antonoglou}}, \bibinfo {author} {\bibfnamefont {H.}~\bibnamefont {King}}, \bibinfo {author} {\bibfnamefont
  {D.}~\bibnamefont {Kumaran}}, \bibinfo {author} {\bibfnamefont {D.}~\bibnamefont {Wierstra}}, \bibinfo {author} {\bibfnamefont {S.}~\bibnamefont {Legg}},\ and\ \bibinfo {author} {\bibfnamefont {D.}~\bibnamefont {Hassabis}},\ }\href {http://dx.doi.org/10.1038/nature14236} {\bibfield  {journal} {\bibinfo  {journal} {Nature}\ }\textbf {\bibinfo {volume} {518}},\ \bibinfo {pages} {529} (\bibinfo {year} {2015})}\BibitemShut {NoStop}%
\bibitem [{\citenamefont {Schulman}\ \emph {et~al.}(2015)\citenamefont {Schulman}, \citenamefont {Levine}, \citenamefont {Abbeel}, \citenamefont {Jordan},\ and\ \citenamefont {Moritz}}]{schulman2015trust}%
  \BibitemOpen
  \bibfield  {author} {\bibinfo {author} {\bibfnamefont {J.}~\bibnamefont {Schulman}}, \bibinfo {author} {\bibfnamefont {S.}~\bibnamefont {Levine}}, \bibinfo {author} {\bibfnamefont {P.}~\bibnamefont {Abbeel}}, \bibinfo {author} {\bibfnamefont {M.}~\bibnamefont {Jordan}},\ and\ \bibinfo {author} {\bibfnamefont {P.}~\bibnamefont {Moritz}},\ }in\ \href {https://proceedings.mlr.press/v37/schulman15.html} {\emph {\bibinfo {booktitle} {Proceedings of the 32nd International Conference on Machine Learning}}},\ \bibinfo {series} {Proceedings of Machine Learning Research}, Vol.~\bibinfo {volume} {37},\ \bibinfo {editor} {edited by\ \bibinfo {editor} {\bibfnamefont {F.}~\bibnamefont {Bach}}\ and\ \bibinfo {editor} {\bibfnamefont {D.}~\bibnamefont {Blei}}}\ (\bibinfo  {publisher} {PMLR},\ \bibinfo {address} {Lille, France},\ \bibinfo {year} {2015})\ pp.\ \bibinfo {pages} {1889--1897}\BibitemShut {NoStop}%
\bibitem [{\citenamefont {Brockman}\ \emph {et~al.}(2016)\citenamefont {Brockman}, \citenamefont {Cheung}, \citenamefont {Pettersson}, \citenamefont {Schneider}, \citenamefont {Schulman}, \citenamefont {Tang},\ and\ \citenamefont {Zaremba}}]{brockman2016openaigym}%
  \BibitemOpen
  \bibfield  {author} {\bibinfo {author} {\bibfnamefont {G.}~\bibnamefont {Brockman}}, \bibinfo {author} {\bibfnamefont {V.}~\bibnamefont {Cheung}}, \bibinfo {author} {\bibfnamefont {L.}~\bibnamefont {Pettersson}}, \bibinfo {author} {\bibfnamefont {J.}~\bibnamefont {Schneider}}, \bibinfo {author} {\bibfnamefont {J.}~\bibnamefont {Schulman}}, \bibinfo {author} {\bibfnamefont {J.}~\bibnamefont {Tang}},\ and\ \bibinfo {author} {\bibfnamefont {W.}~\bibnamefont {Zaremba}},\ }\href {https://arxiv.org/abs/1606.01540} {\bibinfo {title} {Openai gym}} (\bibinfo {year} {2016}),\ \Eprint {https://arxiv.org/abs/1606.01540} {arXiv:1606.01540 [cs.LG]} \BibitemShut {NoStop}%
\bibitem [{Note1()}]{Note1}%
  \BibitemOpen
  \bibinfo {note} {Odd and even, basically the same. On average, one needs at least $0.5$ more measurement when the number of qubits increased by one.}\BibitemShut {Stop}%
\bibitem [{Note2()}]{Note2}%
  \BibitemOpen
  \bibinfo {note} {We leave detailed analytical studies of the behavior from the aspect of entanglement membranes and the percolation theory for future work.}\BibitemShut {Stop}%
\bibitem [{\citenamefont {Hu}\ \emph {et~al.}(2024)\citenamefont {Hu}, \citenamefont {Zhao}, \citenamefont {Patti}, \citenamefont {Gu}, \citenamefont {Gomez}, \citenamefont {Abney-McPeek}, \citenamefont {You},\ and\ \citenamefont {Yelin}}]{Hu2024}%
  \BibitemOpen
  \bibfield  {author} {\bibinfo {author} {\bibfnamefont {H.-Y.}\ \bibnamefont {Hu}}, \bibinfo {author} {\bibfnamefont {C.}~\bibnamefont {Zhao}}, \bibinfo {author} {\bibfnamefont {T.~L.}\ \bibnamefont {Patti}}, \bibinfo {author} {\bibfnamefont {A.}~\bibnamefont {Gu}}, \bibinfo {author} {\bibfnamefont {A.~M.}\ \bibnamefont {Gomez}}, \bibinfo {author} {\bibfnamefont {F.}~\bibnamefont {Abney-McPeek}}, \bibinfo {author} {\bibfnamefont {Y.-Z.}\ \bibnamefont {You}},\ and\ \bibinfo {author} {\bibfnamefont {S.~F.}\ \bibnamefont {Yelin}},\ }\href {https://github.com/hongyehu/PyClifford} {\bibfield  {journal} {\bibinfo  {journal} {Manuscript in preparation}\ } (\bibinfo {year} {2024})},\ \bibinfo {note} {unpublished manuscript, available upon request mailto:hongyehu@g.harvard.edu}\BibitemShut {NoStop}%
\bibitem [{\citenamefont {Raffin}\ \emph {et~al.}(2021)\citenamefont {Raffin}, \citenamefont {Hill}, \citenamefont {Gleave}, \citenamefont {Kanervisto}, \citenamefont {Ernestus},\ and\ \citenamefont {Dormann}}]{stable-baselines3}%
  \BibitemOpen
  \bibfield  {author} {\bibinfo {author} {\bibfnamefont {A.}~\bibnamefont {Raffin}}, \bibinfo {author} {\bibfnamefont {A.}~\bibnamefont {Hill}}, \bibinfo {author} {\bibfnamefont {A.}~\bibnamefont {Gleave}}, \bibinfo {author} {\bibfnamefont {A.}~\bibnamefont {Kanervisto}}, \bibinfo {author} {\bibfnamefont {M.}~\bibnamefont {Ernestus}},\ and\ \bibinfo {author} {\bibfnamefont {N.}~\bibnamefont {Dormann}},\ }\href {http://jmlr.org/papers/v22/20-1364.html} {\bibfield  {journal} {\bibinfo  {journal} {Journal of Machine Learning Research}\ }\textbf {\bibinfo {volume} {22}},\ \bibinfo {pages} {1} (\bibinfo {year} {2021})}\BibitemShut {NoStop}%
\end{thebibliography}%

\clearpage
\onecolumngrid
\appendix

\section{Numerical Methods}
\label{app:num-methods}

In this section, we provide details on the numerical methods used to simulate efficient disentanglers. Our methods heavily rely on Clifford circuits and stabilizer states, for that end the sampling of random 2-qubit Clifford gates and von Neumann entropy calculations using stabilizer states were performed using the PyClifford package \cite{Hu2024}. As discussed in the main text, we have used proximal policy optimization methods to find the reinforced disentanglers. For that end, the discrete observation and action spaces were constructed using the Open AI Gym package \cite{brockman2016openaigym}, while the proximal policy optimization model was initiated, trained, and saved using Stable-Baselines3 \cite{stable-baselines3}.

Depending on the circuit size, we have trained our models using $t_s = 2\times 10^5, 2.5\times 10^5, 5\times 10^5$ maximum time steps, an entropy coefficient of $e_c = 10^{-2}$, and a learning rate of $l_r = 10^{-3}$. 

\begin{figure}[!h]
    \centering
    \includegraphics[width=\linewidth]{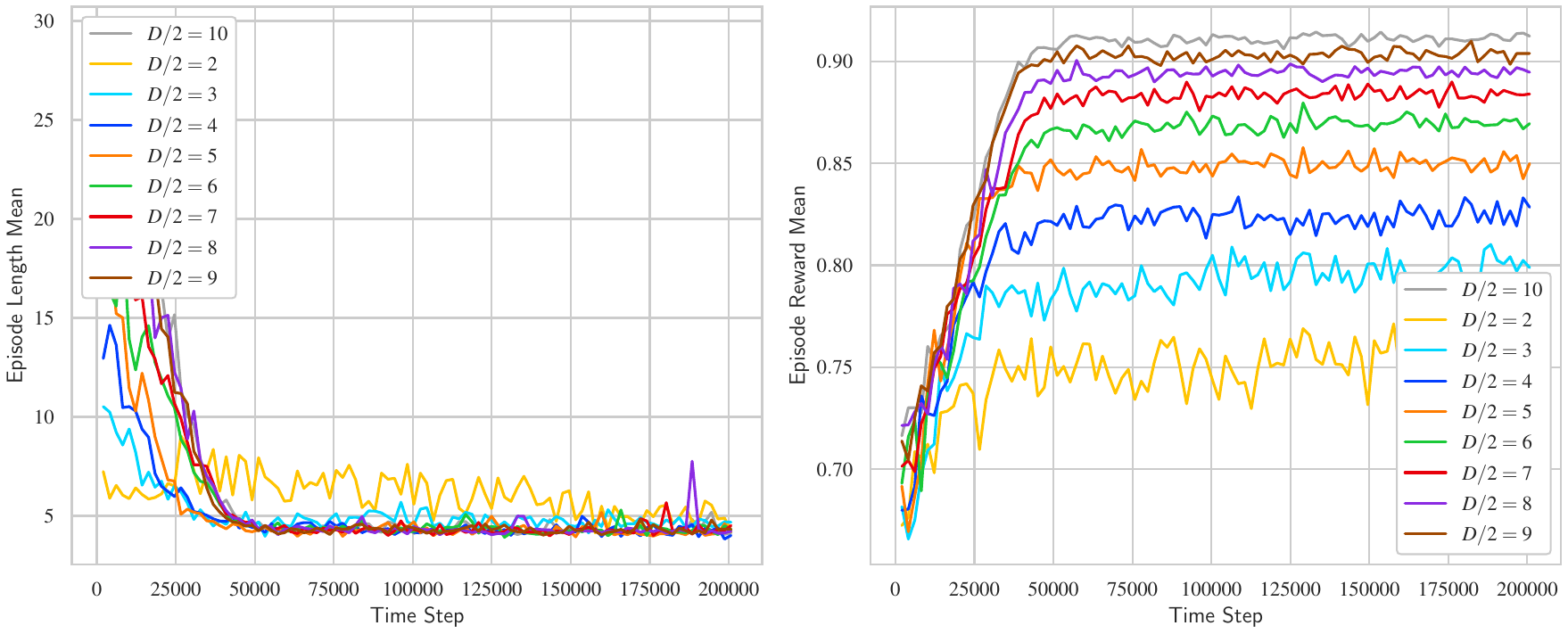}
    \caption{Mean episode length and mean episode reward as a function of time for increasing depth. $N = 6$ and $\alpha = 0.1$.}
    \label{fig:tb_n_6}
\end{figure}

For circuit sizes $N = [3,4,5,6,7,8,9,10,11]$, $D/2 = [2,3,4,5,6,7,8,9,10,11]$ we have fixed $\alpha=0.1$ and trained for $t_s = 200,000$. Metrics for the learning over time can be found in figure \ref{fig:tb_n_6}.
For these simulations, we have resorted to a sparse reward function, such that the agent is rewarded only when it succeeds, and the cost of measurements is subtracted from the reward. At intermediate steps where the entropy is non-zero, the reward is fixed to zero.

\begin{table}[!h]
    \centering
    \begin{tabular}{|c|c|c|c|c|c|c|c|c|c|}
        \hline Rewards & $N$   & $\gamma_1$            & $\gamma_2$           & $\gamma_3$      & Measurements & $N$   & $\gamma_1$        & $\gamma_2$         & $\gamma_3$ \\
        \hline         & 3     & 0.199631       & 0.234615      & 0.744301 &              & 3     & 0.69849    & 0.2617      & 0.683677 \\
        \hline         & 4     & -0.284870       & -0.2755       & 0.642525 &              & 4     & 1.089911   & 0.072598    & 1.568935 \\
        \hline         & 5     & 0.176500         & 0.160431      & 0.755519 &              & 5     & 2.608932   & 0.261157    & 0.626993 \\
        \hline         & 6     & 0.255514       & 0.181856      & 0.666857 &              & 6     & 3.578105   & 0.369607    & 0.624915 \\
        \hline         & 7     & 0.221658       & 0.188698      & 0.693190  &              & 7     & 3.639860    & 0.236387    & 1.548429 \\
        \hline         & 8     & 0.285976       & 0.189250       & 0.625591 &              & 8     & 4.741684   & 0.338947    & 1.472527 \\
        \hline         & 9     & 0.260782       & 0.199553      & 0.647152 &              & 9     & 4.560935   & 0.247167    & 2.456257 \\
        \hline         & 10    & 0.327017       & 0.206863      & 0.575002 &              & 10    & 5.345993   & 0.316333    & 2.876648 \\
        \hline Rewards & $D/2$ & $\gamma_1$            & $\gamma_2$           &          & Measurements & $D/2$ & $\gamma_1$        & $\gamma_2$         & \\
        \hline         & 2	   & -0.01584903	& 0.86759581	&          &              &	2	  & 0.65911905 & -1.12977381 & \\ 
        \hline         & 3	   & -0.0142489	    & 0.90259737	&          &              &	3	  & 0.7920119  & -1.40795235 & \\
        \hline         & 4	   & -0.01204827	& 0.91725471	&          &              &	4	  & 0.8650119  & -1.50420238 & \\
        \hline         & 5	   & -0.01059102	& 0.92935558	&          &              &	5	  & 0.9120000  & -1.57749999 & \\
        \hline         & 6	   & -0.00951192	& 0.93983818	&          &              &	6	  & 0.93965476 & -1.65563095 & \\
        \hline         & 7	   & -0.00867545	& 0.94648441	&          &              &	7	  & 0.97002381 & -1.74515476 & \\
        \hline         & 8	   & -0.00715495	& 0.9458137		&          &              &  8	  & 0.94880952 & -1.55026190 & \\
        \hline         & 9	   & -0.0069116	    & 0.9516103		&          &              &  9	  & 0.97405952 & -1.64026191 & \\
        \hline         & 10	   & -0.00634318	& 0.95471422	&          &              &	10	  & 0.96753572 & -1.59435715 & \\
        \hline
        \end{tabular}
    \caption{Best-fit coefficients for $\tanh$ and linear fit for depth and number of qubit dependence of rewards and number of projections averaged over episodes for the trained PPO model.}
    \label{tab:my_label}
\end{table}

\begin{figure}[!h]
    \centering
    \includegraphics[width=\linewidth]{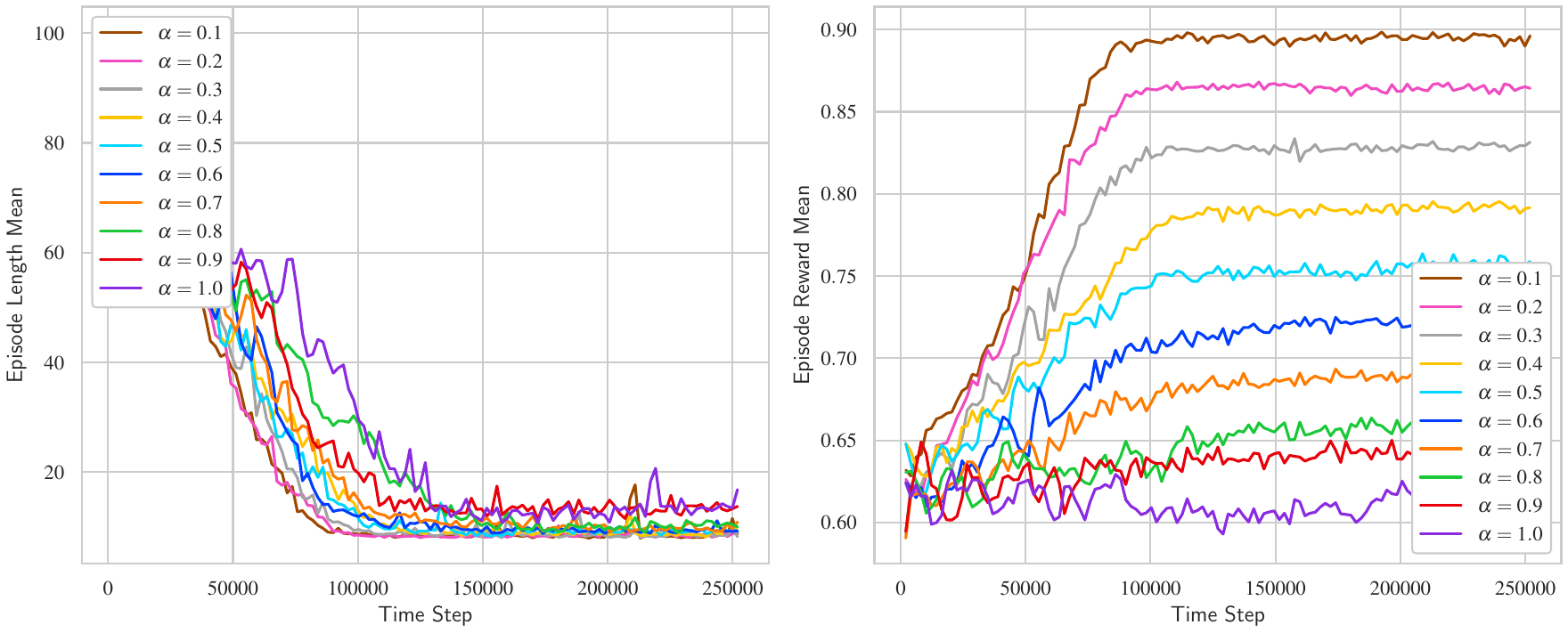}
    \caption{Mean episode length and mean episode reward as a function of time for increasing values of penalty slope. Circuit size $N\times D = 10 \times 20$.}
    \label{fig:tb_alpha_10x10}
\end{figure}

For fixed circuit sizes $N,D/2 = [5\times 5, 10\times 10, 15\times 15]$ we have also simulated efficient disentanglers for varying penalty slopes $\alpha = 0.1,0.2,0.3,0.4,0.5,0.6,0.7,0.8,0.9,1.0$. Metrics for the learning over time can be found in figure  \ref{fig:tb_alpha_10x10}.





\end{document}